\documentstyle[preprint,aps,a4,12pt,axodraw,psfig]{revtex}

\newcommand{\pr}{\hspace{\parindent}}
\begin{document}
\draft
\hfill\vbox{\baselineskip14pt
            \hbox{\bf KEK-TH-490}
            \hbox{KEK Preprint 96-87}
            \hbox{MPI-PhT/96-79}
            \hbox{hep-ph/9607466}
            \hbox{Revised Version}}
\baselineskip20pt
\vskip 0.2cm 
\begin{center}
{\Large\bf One Loop Supersymmetric QCD Radiative Corrections
  to the top quark production in
   $p\bar{p}$ collisions}
\end{center} 
\vskip 0.2cm 
\begin{center}
\large S.Alam$^{1}$, K. Hagiwara$^{1,2}$, and S. Matsumoto$^{1}$ 
\end{center}
\begin{center}
{\it $^{1}$Theory Group, KEK, Tsukuba, Ibaraki 305, Japan\\
     $^{2}$Max-Planck-Institut f\"ur Physik,\\
         F\"ohringer Ring 6, 80805 M\"unchen, Germany}
\end{center}
\begin{center} 
\large Abstract
\end{center}
\begin{center}
\begin{minipage}{14cm}
\baselineskip=15pt
\noindent
The purpose of this note is to give the one loop radiative
corrections to the top quark pair production in the $p\bar{p}$
annihilation at the Fermilab Tevatron in the context of the
Minimal Supersymmetric Model. We concentrate here on the
supersymmetric QCD corrections and give the analytic expression
for these corrections. Recently Li et. al. have reported the
supersymmetric QCD corrections to this process we indicate 
clearly a comparison of their and our work. In particular, we
find additional corrections [crossed box and gluon self-energy]
at the one loop level which are not given by Li et. al.. 
Our numerical results disagree with the original claim of Li et. al.
The numerical values given by them in a recent erratum do agree
with the general trend of our numerical results however 
the actual values still disagree. We find that the percentage
corrections at the hadronic corrections changes from $22\%$
to $-0.5\%$ as the squark mass is changed from 100 GeV to 600 GeV,
for a gluino mass of 200 GeV. For a  gluino mass of 150 GeV
the squark-mass dependence is less abrupt; they change from
$-5.3\%$ to $1\%$ as the  squark mass is varied between 100 GeV
and 600 GeV. We also present numerical results for differential
cross section at the hadronic level, and percentage corrections
at the parton level.


\end{minipage}
\end{center}
\vfill

\baselineskip=20pt
\normalsize

\newpage
\setcounter{page}{2}
\section{Introduction}
\pr

As is well known by now the top quark existence has been experimentally
shown by the CDF \cite{Abe95}and D0 \cite{Aba95} 
at almost $100\%$ confidence level.
Two interesting parameters, the mass of top and the cross section
for top pair production, have been found as follows:
 by the CDF \cite{Abe95}
\begin{enumerate}
\item{\hspace{ 5mm}}$\rm{m_{t}^{expt.}=176\pm 9} $ GeV,
\item{\hskip 5mm}$\rm{\sigma_{t\bar{t}}^{expt.}=7.6^{+1.9}_{-1.5}}$ pb.
 \end{enumerate}
The D0 \cite{Aba95} finds for the same parameters 
\begin{enumerate}
\item{\hspace{ 5mm}}$\rm{m_{t}^{expt.}=170\pm 18} $ GeV,
\item{\hskip 5mm }$\rm{\sigma_{t\bar{t}}^{expt.}=5.2\pm 1.8}$ pb.
 \end{enumerate}
The standard model theoretical predictions for the top pair production
cross section is, assuming a top mass of 170 and 175 GeV \cite{Ber95},
\begin{enumerate}
\item{\hskip 5mm }\rm{$\sigma_{t\bar{t}}^{theory}=
6.48^{+0.09}_{-0.48}$ pb, $m_t=170$} GeV,
\item{\hskip 5mm }\rm{$\sigma_{t\bar{t}}^{theory}=
5.52^{+0.07}_{-0.42}$ pb, $m_t=175$} GeV.
 \end{enumerate}
A theoretical fit based on the Standard Model Electroweak Precision
calculations gives for the top mass the following limits \cite{Hag96}
\begin{equation}
  \rm{   m_t=179\pm 7^{+19(m_H=1000\; GeV)}_{-22(m_H=60\; GeV)}
\mp 2(\alpha_s)\mp 5(\delta_{\alpha})}\nonumber
\end{equation} 
here $\alpha_s=0.120\pm 0.07$ and $\delta_{\alpha}=0.03\pm 0.09$
are the uncertainities in ${\rm \alpha_{s}(m_Z)}$ \cite{Pdg94} and 
${\rm \alpha(m_Z^2)}$ \cite{Eid95}, respectively.

	Once the main injector upgrade becomes operational 
in 1999 \cite{Ami96}
at Fermilab, the experimental sensitivity will be 
highly increased. For example
the uncertainty in the production cross section will be reduced to
6-11$\%$. The top mass uncertainty will be reduced to around 1-2$\%$.
Clearly the agreement between standard model theory and experimental
results is not close enough to include moderate shifts from the SM
results. 	

	We have considered the
complete one loop SUSY corrections to the process 
$q\bar{q}\longrightarrow t\bar{t}$. These include SUSY-QCD
and SUSY-QFD corrections not ignoring the box [both direct
and crossed boxes]. Although box diagrams in general give a
small contributions
one must include them for completeness and exact numerical
predictions. The purpose of this note is to concentrate
on the complete one loop supersymmetric QCD corrections.  
Recently Li et. al.\cite{Li95} have reported the one
loop SUSY-QCD corrections. We give a comparison between their
and our work. In particular we find additional corrections
[crossed box and gluon self-energy] which are not given by them.
Several mistakes/misprints in their work are also noted, however
their erratum \cite{Li95} now corrects these. Importantly our
numerical work does not agree with their original claim \cite{Li95}.
However their numerical values given in the erratum \cite{Li95}
agrees now with the general trend that we give in this report.
 There still remains some disagreements as we report in section 4.
We also give the energy dependence of the cross section and differential
cross sections at the parton level. This allows us to 
compare our results with the one loop correction 
to the same
subprocess in the standard model \cite{Been94}. 
One may use the parton level values to study the details of 
the SUSY corrections more directly. 

	The one loop Electroweak corrections to the process 
$\rm{q\bar{q}\longrightarrow t\bar{t}}$ have been considered
by several groups\cite{sta93}in the context of MSSM. 
Our results on these  will be be presented
in a subsequent paper. The complete SUSY corrections 
to the processes
$\rm{q\bar{q}\longrightarrow q \bar{q}}$, $\rm{qq\longrightarrow qq}$,
$\rm{q\bar{q}\longrightarrow g g }$, and $\rm{qg\longrightarrow qg}$
are being considered by \cite{John96}. 
 
The layout of this paper is as follows. In next section we give the
one loop SQCD radiative corrections to the process 
$\rm{q\bar{q}\longrightarrow t\bar{t}}$
which arise from the gluon self-energy, the quark wave function 
renormalization and the triangle diagrams.
For completeness we also include the Born expression for the process
$\rm{q\bar{q}\longrightarrow t \bar{t}}$.
In Sec.~3, we write out the results for the corrections arising from
the box diagrams [direct and crossed] due to the squarks and gluinos.
Sec.~4 gives the numerical results. For our
numerical work we use the Fortran code FF \cite{Old90}for the 
evaluation of  
the scalar integrals \cite{Pas79}, and 
the MRSA parton distributions of Martin et. al.
\cite{Mar94} and finally the integrations  are 
carried out by using BASES \cite{Kaw95}. We have made several cross
checks to make sure to eliminate any numerical errors. In the
appendix we give the box contribution using the same momentum 
assignment of \cite{Li95}.
We compare our results with \cite{Li95} wherever required.

\section{Tree, and the one-loop contributions in SQCD [except for box ] 
         to the process \lowercase{$Q\bar{Q}
\longrightarrow T\bar{T}$}.}
\pr
	At the parton level the processes responsible for the production 
of top[t] anti-top[$\bar{t}$]in energetic
$p\bar{p}$ collisions to order[$\alpha_s^2$ i.e. tree-level] 
are
\begin{itemize}
\item{\hskip 1.0 cm}The annihilation of quark-antiquark pair into
top anti-top via a virtual gluon exchange 
\begin{center}${\rm q[p_1]\bar{q}[p_2]
\longrightarrow t [p_3]\bar{t}[p_4]}$
\end{center}
\item{\hskip 1.0 cm}The fusion of gluon-pair into
top anti-top via a virtual gluon or a virtual top-quark exchange\\ 
\begin{center}
${\rm g[p_1]g[p_2]
\longrightarrow t [p_3]\bar{t}[p_4]}$
\end{center}
\end{itemize}
Particle momenta have been shown in the parentheses.
The schematic diagram for the first process is shown in Fig. 1, which
is the reaction we choose to concentrate in this paper. To 
get a complete
analysis one must include the second process as is done 
for the standard
model \cite{Been94}, although it contributes only $10\%$ at the 
Tevatron. We work with the Mandelstam variables~ s,~t, and u defined as
	\begin{eqnarray}
	\rm{\hat{s} = (p_1+p_2)^2 = (p_3+p_4)^2}, \\	
	\rm{\hat{t} = (p_1-p_3)^2 = (p_4-p_2)^2}, \\	
	\rm{\hat{u} = (p_2-p_3)^2 = (p_4-p_1)^2} 	
	\end{eqnarray}
The Mandlestam variables satisfy the  
relation ${\rm \hat{s}+\hat{t}+\hat{u}=2m_t^2}$
where we have taken the initial parton mass as zero.
 With our momentum assignments the leading order QCD matrix element
of quark antiquark annihilation is given by
\begin{equation}
{\rm i M^{q\bar{q}}_{Born}=
\bar{u}^{j}_t(p_3,s_3)[-ig_sT^c_{jl}\gamma_{\mu}]v^l_{\bar{t}}(p_4,s_4)
[-i\frac{g^{\mu\nu}}{\hat{s}}]
\bar{v}^{k}_{\bar{q}}(p_2,s_2)[-ig_sT^c_{ki}\gamma_{\nu}]
u^i_{q}(p_1,s_1)}
\end{equation}
It is straightforward to obtain from the above equation the
square of the Born matrix element averaged over initial spin
and color degrees of freedom and summed over the final ones.
We immediately obtain
\begin{equation}
\bar{\sum}|M^{q\bar{q}}_{Born}|=|M_0|^2=
\frac{4g_s^4}{9\hat{s}^2}F_{1}
\end{equation}
here and elsewhere in this paper we define 
\begin{equation}
\rm{F_{1}=2\hat{s}m_t^2+(\hat{t}-m_t^2)^2+(\hat{u}-m_t^2)^2}
\end{equation}
The Born differential cross section is readily written as    
\begin{equation}
\frac{d\sigma_0^{Born}}{d\hat{t}}=\frac{1}{16\pi\hat{s}^2}|M_0|^2
\end{equation}
Using the above equation to integrate over $\hat{t}$, and noting that
the integration limits of $\hat{t}$ are in our case are given by 
the equation
\begin{equation}
{\rm m_t^2-\frac{1}{2}\hat{s}-\frac{1}{2}\hat{s}\beta_t \leq \hat{t}
\leq m_t^2-\frac{1}{2}\hat{s}+\frac{1}{2}\hat{s}\beta_t} 
\end{equation}
[here ${\rm \beta_t=\sqrt{1-\frac{4m_t^2}{\hat{s}}}}$] 
one immediately has
for the expression for the Born cross section at the parton level
\begin{equation}
{\rm \sigma_0^{Born}=\frac{4\pi\alpha_{s}^2}{3 \hat{s}^2}[\frac{2}{9}]
\beta_t[\hat{s}+2m_t^2]}
\end{equation}
We have intentionally written the above result in the form with the
color factor separated out.
	The total amplitude squared upto one loop can be written
as
 \begin{equation}
\rm{|M_{0+1}|^2=|M_0|^2+2 Re [M_{sew}M_0^{\dagger}
+M_{triangle}M_0^{\dagger}
+M_{box}M_0^{\dagger}]}
\end{equation}
 	The total self-energy and wave-function renormalization [sew]
contribution to the process $q\bar{q}\longrightarrow
t\bar{t}$ can be written as
\begin{equation}
\rm{M_{sew}M_0^{\dagger}=M_{SQCD}^{gluon}M_0^{\dagger}
+M_{SQCD}^{w}M_0^{\dagger}}
\end{equation}
	The gluon self energy diagram is shown in Fig. 2a. 
We renormalize the SUSY contribution to the QCD coupling $\alpha_{s}$
at zero momentum transfer.
The gluon self-energy gets contribution from a gluino loop and s-quark
loop. One may write
\begin{equation}
\rm{M^{gluon}_{SQCD}M_0^{\dagger}=M^{gg}_{\tilde {g}}M_0^{\dagger}
+M^{gg}_{\tilde{q}}M_0^{\dagger}}
\end{equation}
\begin{equation}
\rm{M^{gg}_{\tilde {g}}M_0^{\dagger}=
|M_0|^2\frac{\alpha_s}{\pi}(3)\int_0^1
x(1-x)\ln[1-x(1-x)\frac{\hat{s}}{m_{\tilde g}^2}]}
\end{equation}
\begin{equation}
\rm{M^{gg}_{\tilde {q}}M_0^{\dagger}=
|M_0|^2\frac{\alpha_s}{4\pi}(\frac{1}{2})
\int_0^1 \sum_{flavor}
(1-2x)^2\ln[1-x(1-x)\frac{\hat{s}}{m_{\tilde q_1}^2}]}
\end{equation}
The sum over two complex scalar fermions \cite{Hab85}, $\tilde{f}_1$ and 
$\tilde{f}_2$, for each flavor is understood.  For light quarks, 
the mass eigenstates are expected not to deviate significantly 
from the current eigenstates, $\tilde{f}_L$ and $\tilde{f}_R$, 
due to chiral invariance.  This may not be the case for 
the bottom and top squarks.
To avoid plethora of indices we write all our results
for the s-particle 1. However one must be careful that 
the total results do not always follow by replacing 1 by 2, 
for example in the box diagram we can have the mixed case i.e we 
may have squark [mass eigenstate 1] on one side and 
stop [mass eigenstate 2] on the other side of the box.
We denote the squark mixing angle by $\tilde{\theta}$ and that of stop
by $\theta$. The expressions for squark and stop mass eigenstates are
\begin{equation}
{\rm \tilde{q}_1=\tilde{q}_{L}\cos\tilde{\theta}
+\tilde{q}_{R}\sin\tilde{\theta}}
\end{equation}
\begin{equation}
{\rm \tilde{q}_2=-\tilde{q}_{L}\sin\tilde{\theta}
+\tilde{q}_{R}\cos\tilde{\theta}}
\end{equation}
\begin{equation}
{\rm \tilde{t}_1=\tilde{t}_{L}\cos\theta+\tilde{t}_{R}\sin\theta}
\end{equation}
\begin{equation}
{\rm \tilde{t}_2=-\tilde{t}_{L}\sin\theta+\tilde{t}_{R}\cos\theta}
\end{equation}

      The wavefunction renormalization [Fig. 2 a-d] contribution 
can be written as
sum of two terms 
\begin{equation}
\rm{M_{SQCD}^{w}M_0^{\dagger}=M^{wt}M_0^{\dagger}
+M^{wq}M_0^{\dagger}}
\end{equation}
the [wt] contribution comes from the top wave-function renormalization
and [wq] is from the parton [quark] wave function.

\begin{equation}
\rm{M^{wt}M_0^{\dagger}=|M_0|^2\frac{\alpha_s}{3\pi}[A_t^2
(B_1(m_t^2,m_{\tilde{g}},m_{\tilde{t}})+2m_t^2 B_1^{'})
-B_t^2(2 m_{\tilde{g}}m_{t}B_0^{'})]}
\end{equation}

\begin{equation}
\rm{A_t^2 = a_{1}^2+b_{1}^2}
\end{equation}
\begin{equation}
\rm{B_t^2 = a_{1}^2-b_{1}^2}
\end{equation}
\begin{eqnarray*}
a_1=\frac{1}{\sqrt{2}}(\cos\theta-\sin\theta)
\end{eqnarray*}
\begin{eqnarray*}
b_1=\frac{1}{\sqrt{2}}(\cos\theta+\sin\theta)
\end{eqnarray*}
\begin{equation}
\rm{M^{wq}M_0^{\dagger}=|M_0|^2\frac{\alpha_s}{3\pi}[A_q^2
(B_1(0,m_{\tilde{g}},m_{\tilde{q}}))]}
\end{equation}

\begin{equation}
\rm{A_q^2 = \tilde{a}_{1}^2+\tilde{b}_{1}^2}
\end{equation}
and $B_q$ to be used below is given by
\begin{equation}
\rm{B_q^2 = \tilde{a}_{1}^2-\tilde{b}_{1}^2}
\end{equation}
\begin{eqnarray*}
\tilde{a}_1=\frac{1}{\sqrt{2}}(\cos\tilde{\theta}-\sin\tilde{\theta})
\end{eqnarray*}
\begin{eqnarray*}
\tilde{b}_1=\frac{1}{\sqrt{2}}(\cos\tilde{\theta}+\sin\tilde{\theta})
\end{eqnarray*}
 	The total Triangle contribution to the process 
$q\bar{q}\longrightarrow t\bar{t}$ can be written as
\begin{equation}
\rm{M_{triangle}M_0^{\dagger}=M_{T1}M_0^{\dagger}+M_{T2}M_0^{\dagger}
+M_{T3}M_0^{\dagger}+M_{T4}M_0^{\dagger}}
\end{equation}
where [Fig. 3a]
\begin{equation}
\rm{M_{T1}M_0^{\dagger}=\frac{4g_s^4}{9\hat{s}^2}\sum_{i} F_{i}^{T1}}
\end{equation}
Here i=1,2
\begin{eqnarray}
F_{1}^{T1}&=&\frac{\alpha_s}{24\pi}
[2m_{\tilde g}m_t(B_t^2)[C_{0}+C_{11}]
 -2m_t^2 (A_t^2)[C_{21}+C_{11}]-2(A_t^2) C_{24}]F_1
\label{T1a}
\end{eqnarray}

\begin{eqnarray}
F_{2}^{T1}&=&\frac{\alpha_s}{24\pi}
[-2m_{\tilde g}m_t(B_t^2)[C_{0}+C_{11}]
 +2m_t^2 (A_t^2)[C_{21}+C_{11}]]F_2
\nonumber\\
{\rm F_2} &=& {\rm \hat{s}^2}
\label{T1b}
\end{eqnarray}
We note that Li et. al.\cite{Li95} have 
written $\rm{2\hat{s}m_t^2+\hat{s}^2+\hat{s}(\hat{s}-2m_t^2)}$ 
as the coefficient of their ${\rm F_5}$ which is equal to $\rm{2\hat{s}^2}$.
In the above Eqs.\ref{T1a} and \ref{T1b} the arguments of the 
C integral are
${\rm C_{ij}(-p_3,p_5,m_{\tilde{g}},
m_{\tilde{t_1}},m_{\tilde{t_1}})}$.

	The expression for ${\rm M_{T2}M_0^{\dagger}}$ [Fig. 3b] 
is rather simple. We can simply obtain it from the above by 
setting $m_t=0$. As a double check we have also calculated it directly 
and the result reads
\begin{equation}
{\rm M_{T2}M_0^{\dagger}=
\frac{4g_s^4}{9\hat{s}^2}\sum_{i} F_{i}^{T2}}
\end{equation}
Here only i=1 case is nonzero

\begin{eqnarray}
F_{1}^{T2}&=&\frac{\alpha_s}{24\pi}[
-2(A_q^2) C_{24}]F_1
\label{T2a}
\end{eqnarray}
In the above Eq.\ref{T2a} the arguments of the C integral are
$C_{ij}(-p_1,p_5,m_{\tilde{g}},m_{\tilde{q_1}},m_{\tilde{q_1}})$.

	We now give the expression for Fig. 3c
\begin{equation}
{\rm M_{T3}M_0^{\dagger}=
\frac{4g_s^4}{9\hat{s}^2}\sum_{i} F_{i}^{T3}}
\end{equation}
Here i=1,2

\begin{eqnarray}
F_{1}^{T3}&=&\frac{\alpha_s}{24\pi}[9]
[-2m_{\tilde g}m_t(B_t^2)[C_{0}+C_{11}]
 + (A_t^2)[[n-2]C_{24}-\hat{s}(C_{23}-C_{22})\nonumber\\
&&{}\hskip 1.0 cm
-m_t^2(C_{0}+C_{21}+2C_{11})
-m_{\tilde g}^2 C_{0}]]F_1
\label{T3a}
\end{eqnarray}

\begin{eqnarray}
F_{2}^{T3}&=&\frac{\alpha_s}{24\pi}
[9][2m_{\tilde g}m_t(B_t^2)C_{11}
 + 2(A_t^2)[m_t^2(C_{11}+C_{21})]]F_2
\label{T3b}
\end{eqnarray}
In the above Eqs.\ref{T3a} and \ref{T3b} the argument 
of the C integral are
$C_{ij}(-p_3,p_5,m_{\tilde{t_1}},m_{\tilde{g}},m_{\tilde{g}})$.

	The triangle diagram for the $qqg$ vertex is calculated directly
and also as double check got from T3, by first replacing
$m_t$ by $m_q$ and then setting the latter equal to zero.
 One obtains
\begin{equation}
\rm{M_{T4}M_0^{\dagger}=
\frac{4g_s^4}{9\hat{s}^2}\sum_{i} F_{i}^{T4}}
\end{equation}
Here i=1
\begin{eqnarray}
F_{1}^{T4}&=&\frac{\alpha_s}{24\pi}[9][
  (A_q^2)[[n-2]C_{24}-\hat{s}(C_{23}-C_{22})
-m_{\tilde g}^2 C_{0}]]F_1
\label{T4a}
\end{eqnarray}
In the above, Eq.\ref{T4a}, the arguments of the C integral are
${\rm C_{ij}(-p_1,p_5,m_{\tilde{q_1}},m_{\tilde{g}},m_{\tilde{g}})}$.

	One can see from the above contributions of self energy
,wave function renormalization  and triangles that they all factor
into something times tree level amplitude except for contributions
from triangle diagrams, Eqs.\ref{T1b} and \ref{T3b}. These arise
since the top mass cannot be ignored! From the arguments
of the above loop integrals we see immediately that they do not depend
on the t-channel variable. From these simple observations one
can see that the integration over t-channel variable for the above
contributions is straightforward. This is not the case for the box
diagrams since the box loop integrals depend explicitly on
the t and u channel variables.


\section{Contribution from the Box Diagrams}
\pr
The total box contribution to the process $q\bar{q}\longrightarrow
t\bar{t}$ can be written as
\begin{equation}
\rm{M_{box}M_0^{\dagger}=M_{box}^{DB}M_0^{\dagger}
+M_{box}^{CB}M_0^{\dagger}}
\end{equation}
where
\begin{equation}
\rm{M_{box}^{DB}M_0^{\dagger}=\frac{7g_s^4}{432\;\hat{s}}\sum_{i} F_{i}^{DB}}
\end{equation}
Here i=0,11,12,13,23,24,25,26 and 27. By using the notation
\begin{equation}
\rm{A_5^+ A_5^+ = [a_1^2+b_1^2][\tilde{a}_1^2+\tilde{b}_1^2]
+4 a_1 b_1 \tilde{a}_1 \tilde{b}_1}
\end{equation}
\begin{equation}
\rm{A_{5x} A_{5x} = [a_1^2+b_1^2][\tilde{a}_1^2+\tilde{b}_1^2]
-4 a_1 b_1 \tilde{a}_1 \tilde{b}_1}
\end{equation}
\begin{equation}
\rm{A_{5}^{+} A_{5x} = [a_1^2-b_1^2][\tilde{a}_1^2+\tilde{b}_1^2]}
\end{equation}
we find for the direct-box diagram [Fig.~4a] contribution:
\begin{eqnarray}
F_{0}^{DB}&=&\frac{\alpha_s}{\pi}[m_{\tilde g}^{2}(2A_{5x}A_{5x})
           [2 \hat{s} m_t^2+2(\hat{t}-m_t^2)^2]]D_{0}
\label{F0}
\end{eqnarray}
\begin{eqnarray}
F_{11}^{DB}&=&\frac{\alpha_s}{\pi}
        [m_{\tilde g}m_t (2A_{5}^{+} A_{5x})
         [-2\hat{s}^2]] D_{11}
\label{F11}
\end{eqnarray}
\begin{eqnarray}
F_{12}^{DB}&=&\frac{\alpha_s}{\pi}[
         m_{\tilde g}m_t (2A_{5}^{+} A_{5x})
         [2\hat{s}(\hat{s}-2m_t^2)-4(\hat{t}-m_t^2)^2]] D_{12}
\label{F12}
\end{eqnarray}
\begin{eqnarray}
F_{13}^{DB}&=&\frac{\alpha_s}{\pi}[
 -m_{\tilde g}m_t (2A_{5}^{+} A_{5x})
[2\hat{s}(\hat{s}-2m_t^2)-4(\hat{t}-m_t^2)^2]] D_{13}
\label{F13}
\end{eqnarray}

\begin{eqnarray}
F_{22}^{DB}&=& \frac{\alpha_s}{\pi}[m_t^2(2A_5^+A_5^+)
[2\hat{s}m_t^2+2(\hat{t}-m_t^2)^2-2\hat{s}^2]]D_{22}
\end{eqnarray}

\begin{eqnarray}
F_{23}^{DB}&=& \frac{\alpha_s}{\pi}[m_t^2(2A_5^+A_5^+)
[2\hat{s}m_t^2+2(\hat{t}-m_t^2)^2-2\hat{s}^2]]D_{23}
\end{eqnarray}

\begin{eqnarray}
F_{24}^{DB}&=& \frac{\alpha_s}{\pi}[m_t^2(2A_5^+A_5^+)
[2\hat{s}^2]]D_{24}
\end{eqnarray}
\begin{eqnarray}
F_{25}^{DB}&=& \frac{\alpha_s}{\pi}[(2A_5^+A_5^+)
[2\hat{s}(\hat{u}-m_t^2)^2]]D_{25}
\end{eqnarray}

\begin{eqnarray}
F_{26}^{DB}&=& \frac{\alpha_s}{\pi}[(2A_5^+A_5^+)
[2\hat{s}^2 m_t^2-2m_t^2(2\hat{s}m_t^2+2(\hat{t}-m_t^2)^2)
-2\hat{s}(\hat{u}-m_t^2)^2]]D_{26}
\end{eqnarray}

\begin{eqnarray}
F_{27}^{DB}&=& \frac{\alpha_s}{\pi}[-2(2A_5^+A_5^+)
[2\hat{s}m_t^2+2(\hat{u}-m_t^2)^2]]D_{27}
\end{eqnarray}
In the above, the arguments of the D-functions are
$\rm{D_i=D_i[-p_1,p_3,p_4,
m_{\tilde {q_1}},m_{\tilde{g}},m_{\tilde{t_1}},
m_{\tilde{g}}]}$.
 	
	For the contribution of the crossed-box diagram [Fig. 4b], we find
\begin{equation}
\rm{M_{box}^{CB}M_0^{\dagger}=[-\frac{2}{7}]
\frac{7g_s^4}{432\;\hat{s}}\sum_{i} F_{i}^{CB}}
\end{equation}
with
\begin{equation}
\rm{\overline{A_{5}^{+} A_{5x}}=-[a_1^2+b_1^2][\tilde{a}_1^2-\tilde{b}_1^2]}
\end{equation}
\begin{eqnarray}
F_{0}^{CB}&=&\frac{\alpha_s}{\pi}[m_{\tilde g}^{2}(2A_{5x}A_{5x})
         [2\hat{s} m_t^2+2(\hat{u}-m_t^2)^2]]D_{0}
\label{F0C}
\end{eqnarray}

\begin{eqnarray}
F_{11}^{CB}&=&\frac{\alpha_s}{\pi}
        [m_{\tilde g}m_t (2\overline{A_{5}^{+} A_{5x}})
         [-2 \hat{s}^2]] D_{11}
\label{F11C}
\end{eqnarray}
\begin{eqnarray}
F_{12}^{CB}&=&\frac{\alpha_s}{\pi}[
       m_{\tilde g}m_t (2\overline{A_{5}^{+} A_{5x}})
  [2 \hat{s}(\hat{s}-2m_t^2)-4(\hat{u}-m_t^2)^2]] D_{12}
\label{F12C}
\end{eqnarray}
\begin{eqnarray}
F_{13}^{CB}&=&\frac{\alpha_s}{\pi}[
 -m_{\tilde g}m_t (2\overline{A_{5}^{+} A_{5x}})
  [2 \hat{s} (\hat{s}-2m_t^2)-4(\hat{u}-m_t^2)^2]] D_{13}
\label{F13C}
\end{eqnarray}

\begin{eqnarray}
F_{22}^{CB}&=& \frac{\alpha_s}{\pi}[m_t^2(2A_5^+A_5^+)
[2\hat{s}m_t^2+2(\hat{u}-m_t^2)^2-2\hat{s}^2]]D_{22}
\end{eqnarray}

\begin{eqnarray}
F_{23}^{CB}&=& \frac{\alpha_s}{\pi}[m_t^2(2A_5^+A_5^+)
[2\hat{s}m_t^2+2(\hat{u}-m_t^2)^2-2\hat{s}^2]]D_{23}
\end{eqnarray}

\begin{eqnarray}
F_{24}^{CB}&=& \frac{\alpha_s}{\pi}[m_t^2(2A_5^+A_5^+)
[2\hat{s}^2]]D_{24}
\end{eqnarray}
\begin{eqnarray}
F_{25}^{CB}&=& \frac{\alpha_s}{\pi}[(2A_5^+A_5^+)
[2\hat{s}(\hat{t}-m_t^2)^2]]D_{25}
\end{eqnarray}

\begin{eqnarray}
F_{26}^{CB}&=& \frac{\alpha_s}{\pi}[(2A_5^+A_5^+)
[2\hat{s}^2 m_t^2-2m_t^2(2\hat{s}m_t^2
+2(\hat{u}-m_t^2)^2)-2\hat{s}(\hat{t}-m_t^2)^2]]D_{26}
\end{eqnarray}

\begin{eqnarray}
F_{27}^{CB}&=& \frac{\alpha_s}{\pi}[-2(2A_5^+A_5^+)
[2\hat{s}m_t^2+2(\hat{t}-m_t^2)^2]]D_{27}
\end{eqnarray}
In the above, the arguments of the D-functions are
$\rm{D_i=D_i[-p_1,p_4,p_3,m_{\tilde {q_1}},m_{\tilde{g}},m_{\tilde{t_1}},
m_{\tilde{g}}]}$.

	We are now in a position to write the expression for top
pair production in proton anti-proton collision by weighing our
expressions for differential cross section and cross section of
the subprocess $q\bar{q}\longrightarrow t\bar{t}$ by the parton
distribution functions and integrating over the parton variables i.e., 
\begin{equation}
{\rm d\sigma(p\bar{p}\rightarrow t\bar{t})=\sum_{i,j}\;
\int_{0}^{1}dx_1 \int_{0}^{1}dx_2
[D_{i/p}(x_1,Q^2)\;D_{j/\bar{p}}(x_2,Q^2)]
d\hat{\sigma}(ij\rightarrow t\bar{t})}
\end{equation}
	Here $d\hat{\sigma}$ represents the subprocess cross section
at c.m. energy square of $\hat{s}=x_1 x_2 s$, where $\sqrt{s}$ is the
c.m. energy of the $p\bar{p}$ system. In our numerical calculation, we 
adopt the MRSA parametrization \cite{Mar94} for effective parton 
distribution evaluated at $Q^2=m_t^2$. In order to show the SUSY radiative
corrections clearly, we use a constant $\alpha_s=0.123$.


\section{Numerical Results and Conclusions}

       To facilitate comparison with work of \cite{Li95} we give
our numerical results for the same parameter values as in \cite{Li95}.
A more detailed numerical work will be given elsewhere.
We thus take $\rm{m_t=170\; GeV}$, and assume no mixing between the
squarks. The mass splitting between the squarks of different flavors
is also ignored \cite{Li95}. The common squark mass is denoted by
$\rm{m_{\tilde{q}}}$. 

	We first consider percentage one loop corrections at the
hadronic cross section as a function of the squark mass. Taking
the gluino mass to be 150 GeV we find the percentage corrections
changes from $-5.3\%$ [${\rm m_{\tilde{q}}=100 GeV}$] to $1\%$ 
[${\rm m_{\tilde{q}}=600 GeV}$], see the solid curve of Fig.~5. 
This does not agree with the original claim of 
Li {\em et~al.}\cite{Li95}., where they find for
gluino mass of 150 GeV, $23\%$ [${\rm m_{\tilde{q}}=100\; GeV}$]
and $5\%$ [${\rm m_{\tilde{q}}=420\; GeV}$]. However, the corrected
version \cite{Li95}[see Erratum]values
of $-6\%$ [${\rm m_{\tilde{q}}=100 \;GeV}$]and 
$4\%$  [${\rm m_{\tilde{q}}=600 \;GeV}$]
for gluino mass of 150 GeV are in more closer agreement 
with our values. The remaining small discrepancy may probably 
be explained since they\cite{Li95} have not included the
gluon self-energy and the crossed box. For gluino mass of 200 GeV
we find, see Fig.~5b,that the corrections change rapidly from
$22\%$ to $-0.5\%$ as squark mass changes from 100 GeV to 600 GeV.
Here again there is no agreement with the original claim of
\cite{Li95} where they had reported a variation of $6.5\%$ to $0\%$ 
for a gluino mass of 200 GeV. The revised [erratum \cite{Li95}]
 values of $31\%$
[squark mass of 100 GeV] and $6\%$ [squark mass 600 GeV] 
for gluino mass of
200 GeV, are still somewhat different from our values. We 
find for the gluino mass
of 200 GeV, the relative corrections of $22\%$, $9\%$, $6\%$,
$3\%$, $1\%$, $-0.5\%$ for  ${\rm m_{\tilde{q}}=100,\; 200,\;
300,\;400,\;500,\;600\; GeV}$ respectively, which should be
compared with  $31\%$, $18\%$, $11\%$,$9\%$, $7\%$, $6\%$ 
of the revised values of \cite{Li95}.

A comment is in order. It can be noticed from our Fig.~5 that the
corrections change sign as the gluino mass is changed from
150 GeV to 200 GeV. As we have assumed a top mass of 170 GeV the
threshold for top pair production is crossed in this region and
hence the sign change and rapid change in magnitude of relative
corrections occur. 

	Next let us consider percentage one loop corrections at the
hadronic differential cross section as a function of 
the squark mass. As an example, we show the case at the ${\rm t\bar{t}}$
center-of-mass scattering angle ${\rm \theta_{cm}}=10^{\circ}$.
Taking the gluino mass to be 150 GeV we find the percentage corrections
changes from $-7.5\%$ [${\rm m_{\tilde{q}}=100 GeV}$] to $2.5\%$ 
[${\rm m_{\tilde{q}}=600 GeV}$], see the solid curve in Fig.~6. 
For gluino mass of 200 GeV we find, see the dashed curve in 
Fig.~6,that the corrections change rapidly from
$22\%$ to $0\%$ as squark mass changes from 100 GeV to 600 GeV.
As remarked before it is only the box loop correction which depends
on the t-channel variable or on {\rm $\theta_{cm}$}. 
If the box corrections are not larger than the other contributions,
one would naively expect the percentage
differential cross section to show only a weak dependence 
on {\rm $\theta_{cm}$}. This
is indeed the case as can be seen by comparing Figs.~5 and 6. 

	It is useful to give the percentage correction at the
parton level since among other things they facilitate a comparison
with correction found in the context of standard model \cite{Been94}.
Moreover the corrections at the parton level can provide more
direct and detailed tests of SUSY corrections.
We first show in Fig.~7 the percentage corrections to the
{\rm $q\bar{q}\longrightarrow t\bar{t}$} cross section as 
functions of the subprocess c.m. energy $\sqrt{\hat{s}}$.
At the parton level the total percentage cross section varies 
between $20\%$ and $-9\%$ as center of mass energy is varied 
between 350 GeV to 1.8 TeV.
	When gluino mass is 150~GeV (the solid line), the correction 
	is positive very near to the threshold ($\sqrt{\hat{s}} < 430$~GeV) 
	but it gets rapidly negative down to -9\% at around 
	$\sqrt{\hat{s}}\approx 800$~GeV.  The resulting cancellation 
	explains the smallness of the correction at the hadronic 
	level; see the solid line at $m_{\tilde{q}}=200$~GeV in Fig.~5.  
	For $m_{\tilde{g}}=200$~GeV (the dashed line), the correction 
	grows from about 7\% near the threshold to 20\% at 
	$\sqrt{\hat{s}}=400$~GeV where the gluino-pair threshold 
	opens.  This large positive correction slighly above the 
	$t\bar{t}$ threshold explains the large positive correction 
	to the hadronic cross section in Figs. 5 and 6.
We may compare these results to the standard model \cite{Been94}
who report on the one loop virtual [electroweak] relative corrections
to parton {\rm $q\bar{q}\longrightarrow t\bar{t}$} cross section
among other things. As they take the top 
mass of 100 GeV and 250 GeV we can't compare our results directly
with theirs. However we can extrapolate from their Figs. 9 and 10
that for a top-quark  mass of 170 GeV one would obtain 
corrections between
350 GeV and 1.8 TeV of around $10\%$ and $-15\%$.

	In order to clarify structure of the SUSY radiative 
	corrections, we show in Figs.~8a and 8b contributions from 
	the gluon self-energy correction (dotted line), the sum of 
	the triangle and quark wave-function corrections (short-dash),
	the direct-box (long-dash) and the crossed-box (dash-dotted) 
	contributions separately.  Fig.~8a is for $m_{\tilde{g}}=150$~GeV 
	and Fig.~8b is for $m_{\tilde{g}}=200$~GeV, while all the squark 
	masses are set to 200~GeV in both cases.  While the gluon 
	self-energy correction grows with energy, the sum of the triangle 
	and quark wave-function corrections more than compensate for 
	the effects. Direct box diagram contribution is positive
	near the threshold and turns to negative at higher energies.  
	The crossed box diagram partially cancells the direct box 
	contribution. Box contributions are generally smaller than 
	the other corrections as expected from the similarity of 
	the corrections to the total and differential cross sections.  
	The gluino-pair threshold effects are evident in all the 
	curves in Fig.~8b.

	We finally consider the percentage corrections
at the parton level taking $\rm{\sqrt{\hat{s}}=600\; GeV}$ and letting 
the squark mass vary between 100 GeV and 600 GeV. For the gluino
mass of 150 GeV we find that the percentage corrections of
differential cross section vary between $-20\%$ and $6.5\%$, [Fig.~9].
For the gluino mass of 200 GeV the corrections vary 
between $-7\%$ and $4\%$,[Fig.~9].
 
{\bf{Note added}}:After the calculation was completed and the present 
paper was being written up: the following works came to our attention:

 1: J. Kim {\em et~al.}\cite{Kim96} 
examine both the SUSY Electroweak and SUSY QCD
like correction. However they do not include box diagrams 
and claim that
the box contributions are small 
citing J. Ellis and D. Ross, \cite{John96}
and P. Kraus and F. Wilczek, \cite{Wil96} works as evidence.
These authors state that their results for SUSY QCD agree
with Ref. \cite{Li95} while those of SUSY Electroweak disagree
with J.~Yang and C.S.~Li\cite{sta93}.

2: J. Ellis and D. Ross \cite{John96} work at the
 parton level considering the 
processes {\rm $q\bar{q}\longrightarrow q\bar{q}$
 $q q\longrightarrow q q$, $q\bar{q}\longrightarrow g g $, and
 $q g \longrightarrow q g$}.
However they do not consider $t\bar{t}$ cross section as 
it requires separate treatment.

3: P. Krauss and F. Wilczek \cite{Wil96} also studied the
SUSY corrections to the quark gluon scattering processes
in the limit of large SUSY particle masses. 


\section*{Acknowledgements}

The authors would like to thank S.Y.~Choi, C.S~ Kim, and R.~ Szapalski
for useful suggestions and helpful comments.
We would like to thank R.~ Rangarajan 
for bringing Li {\em et~al.}\cite{Li95}  to our attention.
The work of SA is supported by COE fellowship of the Japanese
Ministry of Education, Science and Culture [MONBUSHO],
that of KH by the
Grant-in-Aid for Scientific Research from the Japanese Ministry 
of Education, Science and Culture (No.~05228104), and that of SM
is supported by Japan Society for the Promotion of Science (No.~2474). 



\section*{Figures}

\begin{itemize}

\item[{\bf Fig.~1}]Tree-Level Diagram for the process $q\bar{q}\longrightarrow
t\bar{t}$.

\item[{\bf Fig.~2a}]Schematic diagram for the Gluon Self-Energy 
 due to SQCD particles.  

\item[{\bf Fig.~2b}]Schematic diagram for the Quark WFNR due to the
                    SQCD particles.  
\item[{\bf Fig.~2c}]Schematic diagram for the Anti-Quark WFNR due to the 
                    SQCD particles.  
\item[{\bf Fig.~2d}]Schematic diagram for the Top WFNR due to the 
                    SQCD particles.  
\item[{\bf Fig.~2e}]Schematic diagram for the Anti-Top WFNR due to the 
                    SQCD particles.  
\item[{\bf Fig.~3a}]Triangle contribution from two stops and one gluino
                    to the $t\bar{t}g$ vertex.  
\item[{\bf Fig.~3b}]Triangle contribution from two squarks and one gluino
                    to the $q\bar{q}g$ vertex.  
\item[{\bf Fig.~3c}]Triangle contribution from one stop and two gluinos
                    to the $t\bar{t}g$ vertex. 
\item[{\bf Fig.~3d}]Triangle contribution from one squark and two gluinos
                    to the $q\bar{q}g$ vertex.  
 
\item[{\bf Fig.~4a}]Direct Box contribution from one stop, one squark
                    and two gluinos.
\item[{\bf Fig.~4b}]Crossed Box contribution from one stop, one squark
                    and two gluinos.

\item[{\bf Fig.~5}]One-loop percentage relative corrections to the
                    ${\rm t\bar{t}}$ production cross section in
                    ${\rm p\bar{p}}$ collisions at ${\rm \sqrt{s}=1.8}$~TeV
                   as a function of squark mass for a gluino mass of
                   150~GeV (solid line) and 200~GeV (dashed line).
                   The top quark mass is set to 170~GeV.
                   
\item[{\bf Fig.~6}]One-loop percentage relative corrections to the 
                   ${\rm t\bar{t}}$ production differential cross section
                   at ${\rm \theta_{cm}=10^{\circ}}$ 
                    in ${\rm p\bar{p}}$ collisions at ${\rm \sqrt{s}=1.8}$~TeV.
                   The results are shown for ${\rm m_t= 170~GeV}$
                    as functions of the squark mass for a gluino mass of
                   150~GeV (solid line) and 200~GeV (dashed line).

\item[{\bf Fig.~7}]One-loop percentage relative correction to 
                    the parton-level 
                    ${\rm q\bar{q}\longrightarrow t\bar{t}}$ cross section
                   for ${\rm m_t=170~GeV}$ as functions of the 
                   ${\rm q\bar{q}}$ center of mass energy,
                   ${\rm \sqrt{\hat{s}}}$.  
                   The results are shown for  
                   ${\rm m_{\tilde{q}}=200~ GeV}$ 
                   and ${\rm  m_{\tilde{g}}=150~GeV}$
                   (solid line) or 
                   ${\rm m_{\tilde{g}}=200~GeV}$~(dashed line).

\item[{\bf Fig.~8a}]Individual contributions of the one-loop percentage 
		relative correction to the $q\bar{q}\rightarrow t\bar{t}$
		cross section are given as functions of the $q\bar{q}$ 
		c.m. energy $\sqrt{\hat{s}}$.  Gluon self-energy correction 
		(dotted line), the sum of triangle and quark wave-function
		corrections (short-dashed line), the direct-box diagram 
		(long-dashed line) and the crossed-box diagram
		contributions (dash-dotted line) are shown separately.  
		Also shown by the solid line is the total percentage
		relative correction, as given in Fig.~7.  The mass 
		paramters are: $m_t=170$~GeV, $m_{\tilde{q}}=200$~GeV, 
		and $m_{\tilde{g}}=150$~GeV. 

\item[{\bf Fig.~8b}] Same as Fig.~8a except $m_{\tilde{g}}=200$~GeV.

\item[{\bf Fig.~9}]One-loop percentage relative corrections to 
                    the parton-level 
                    ${\rm q\bar{q}\longrightarrow t\bar{t}}$ 
                    differential cross section
                   at ${\rm \theta_{cm}=10^{\circ}}$ and
                   ${\rm \sqrt{\hat{s}}=600~GeV}$. The results are shown
                    for ${\rm m_t=170~GeV}$ as functions of the squark mass
                    for a gluino mass of
                   150~GeV (solid line) and 200~GeV (dashed line).

\end{itemize}

\appendix
\section*{Contribution from the Box Diagrams}
\pr
As already mentioned we give in this appendix the box results in the
notation of Li {\em et~al.}\cite{Li95}, for 
the purposes of exact comparison.
 The total box contribution to the process $q\bar{q}\longrightarrow
t\bar{t}$ can be written as
\begin{equation}
\rm{M_{box}M_0^{\dagger}=M_{box}^{DB}M_0^{\dagger}
+M_{box}^{CB}M_0^{\dagger}}
\label{A1}
\end{equation}
where
\begin{equation}
\rm{M_{box}^{DB}M_0^{\dagger}=\frac{7g_s^4}{432\;\hat{s}}\sum_{i} F_{i}^{DB}}
\end{equation}
Here i=0,11,12,13,23,24,25,26 and 27.
\begin{equation}
\rm{A_5^+ A_5^+ = [a_1^2+b_1^2][\tilde{a}_1^2+\tilde{b}_1^2]
+4 a_1 b_1 \tilde{a}_1 \tilde{b}_1}
\end{equation}
\begin{equation}
\rm{A_{5x} A_{5x} = [a_1^2+b_1^2][\tilde{a}_1^2+\tilde{b}_1^2]
-4 a_1 b_1 \tilde{a}_1 \tilde{b}_1}
\end{equation}
\begin{equation}
\rm{A_{5}^{+} A_{5x} = [a_1^2-b_1^2][\tilde{a}_1^2+\tilde{b}_1^2]}
\end{equation}
We note that the following relations hold between our A's and
$\sigma$'s of \cite{Li95}
\begin{eqnarray}
2 A_5^+ A_5^+ &=& \sigma_{ij}^2+\lambda_{ij}^2 \nonumber\\
2 A_5^+ A_{5x} &=& \sigma_{ij}\sigma'_{ij}+\lambda_{ij}\lambda'_{ij}\nonumber\\
2 A_{5x} A_{5x} &=& \sigma_{ij}^{'2}+\lambda_{ij}^{'2}
\end{eqnarray}
\begin{eqnarray}
F_{0}^{DB}&=&\frac{\alpha_s}{\pi}[m_{\tilde g}^{2}(2A_{5x}A_{5x})
   [2 \hat{s} m_t^2+2(\hat{t}-m_t^2)^2]
      +m_{\tilde g}m_t (2A_{5}^{+} A_{5x})[2\hat{s}^2]]D_{0}
\label{AF0}
\end{eqnarray}
\begin{eqnarray}
F_{11}^{DB}&=&\frac{\alpha_s}{\pi}
        [m_{\tilde g}m_t (2A_{5}^{+} A_{5x})
         [2 \hat{s}^2]] D_{11}
\label{AF11}
\end{eqnarray}
\begin{eqnarray}
F_{12}^{DB}&=&\frac{\alpha_s}{\pi}[-m_t^{2}(2A_{5}^{+}A_{5}^{+})
 [2 \hat{s} m_t^2+2(\hat{u}-m_t^2)^2]
-m_{\tilde g}m_t (2A_{5}^{+} A_{5x})[2 \hat{s}^2]] D_{12}
\label{AF12}
\end{eqnarray}
\begin{eqnarray}
F_{13}^{DB}&=&\frac{\alpha_s}{\pi}[m_t^{2}(2A_{5}^{+}A_{5}^{+})
         [2(\hat{u}-m_t^2)^2]\nonumber\\
&& {}\hskip 1.0 cm -m_{\tilde g}m_t (2A_{5}^{+} A_{5x})
 [2\hat{s} (\hat{s}-2m_t^2)-4(\hat{t}-m_t^2)^2]] D_{13}
\label{AF13}
\end{eqnarray}

\begin{eqnarray}
F_{23}^{DB}&=& \frac{\alpha_s}{\pi}[m_t^2(2A_5^+A_5^+)
[2\hat{s}m_t^2+2(\hat{t}-m_t^2)^2-2\hat{s}^2]]D_{23}
\end{eqnarray}

\begin{eqnarray}
F_{24}^{DB}&=& \frac{\alpha_s}{\pi}[-(2A_5^+A_5^+)
[2\hat{s}^2m_t^2+2\hat{s}(\hat{u}-m_t^2)^2]]D_{24}
\end{eqnarray}
\begin{eqnarray}
F_{25}^{DB}&=& \frac{\alpha_s}{\pi}[(2A_5^+A_5^+)
[2\hat{s}(\hat{u}-m_t^2)^2]]D_{25}
\end{eqnarray}

\begin{eqnarray}
F_{26}^{DB}&=& \frac{\alpha_s}{\pi}[m_t^2(2A_5^+A_5^+)
[2\hat{s}^2]]D_{26}
\end{eqnarray}

\begin{eqnarray}
F_{27}^{DB}&=& \frac{\alpha_s}{\pi}[-2(2A_5^+A_5^+)
[2\hat{s}m_t^2+2(\hat{u}-m_t^2)^2]]D_{27}
\end{eqnarray}
Here the arguments of the D-functions are
$\rm{D_i=D_i[-p_1,-p_2,p_4,m_{\tilde {g}},m_{\tilde{q_i}},m_{\tilde{g}},
m_{\tilde{t_i}}]}$.
The above results for the direct box agree with \cite{Li95} after taking
account of their erratum. 

	Although Ref.~\cite{Li95} does not give the contributions of the 
crossed-box diagram, we give our result in their notation.
\begin{equation}
\rm{M_{box}^{CB}M_0^{\dagger}=[-\frac{2}{7}]
\frac{7g_s^4}{432\;\hat{s}}\sum_{i} F_{i}^{CB}}
\end{equation}
Here i=0,11,12,13,23,24,25,26 and 27.
We note that the crossed box color factor is smaller by a factor
of $\frac{2}{7}$ compared to  the direct box, and it contributes
destructively with the direct-box contribution; see Eq. \ref{A1}. 
\begin{equation}
\rm{\overline{A_{5}^{+} A_{5x}}=
-[a_1^2+b_1^2][\tilde{a}_1^2-\tilde{b}_1^2]}
\end{equation}
\begin{eqnarray}
F_{0}^{CB}&=&\frac{\alpha_s}{\pi}[m_{\tilde g}^{2}(2A_{5x}A_{5x})
[2\hat{s}m_t^2+2(\hat{u}-m_t^2)^2]
+m_{\tilde g}m_t (2\overline{A_{5}^{+} A_{5x}})[2s^2]]D_{0}
\label{AF0C}
\end{eqnarray}
\begin{eqnarray}
F_{11}^{CB}&=&\frac{\alpha_s}{\pi}
        [m_{\tilde g}m_t (2\overline{A_{5}^{+} A_{5x}})
         [2\hat{s}^2]] D_{11}
\label{AF11C}
\end{eqnarray}
\begin{eqnarray}
F_{12}^{CB}&=&\frac{\alpha_s}{\pi}[-m_t^{2}(2A_{5}^{+}A_{5}^{+})
         [2\hat{s}m_t^2+2(\hat{t}-m_t^2)^2]
-m_{\tilde g}m_t (2\overline{A_{5}^{+} A_{5x}})[2\hat{s}^2]] D_{12}
\label{AF12C}
\end{eqnarray}
\begin{eqnarray}
F_{13}^{CB}&=&\frac{\alpha_s}{\pi}[m_t^{2}(2A_{5}^{+}A_{5}^{+})
         [2(\hat{t}-m_t^2)^2]\nonumber\\
&&{}\hskip 1.0 cm
-m_{\tilde g}m_t (2\overline{A_{5}^{+} A_{5x}})
 [2\hat{s}(\hat{s}-2m_t^2)-4(\hat{u}-m_t^2)^2]] D_{13}
\label{AF13C}
\end{eqnarray}

\begin{eqnarray}
F_{23}^{CB}&=& \frac{\alpha_s}{\pi}[m_t^2(2A_5^+A_5^+)
[2\hat{s}m_t^2+2(\hat{u}-m_t^2)^2-2\hat{s}^2]]D_{23}
\end{eqnarray}

\begin{eqnarray}
F_{24}^{CB}&=& \frac{\alpha_s}{\pi}[-(2A_5^+A_5^+)
[2\hat{s}^2m_t^2+2\hat{s}(\hat{t}-m_t^2)^2]]D_{24}
\end{eqnarray}
\begin{eqnarray}
F_{25}^{CB}&=& \frac{\alpha_s}{\pi}[(2A_5^+A_5^+)
[2\hat{s}(\hat{t}-m_t^2)^2]]D_{25}
\end{eqnarray}

\begin{eqnarray}
F_{26}^{CB}&=& \frac{\alpha_s}{\pi}[m_t^2(2A_5^+A_5^+)
[2\hat{s}^2]]D_{26}
\end{eqnarray}

\begin{eqnarray}
F_{27}^{CB}&=& \frac{\alpha_s}{\pi}[-2(2A_5^+A_5^+)
[2\hat{s}m_t^2+2(\hat{t}-m_t^2)^2]]D_{27}
\end{eqnarray}
Here the arguments of the D-functions are
$\rm{D_i=D_i[-p_1,-p_2,p_3,m_{\tilde {g}},m_{\tilde{q_i}},m_{\tilde{g}},
m_{\tilde{t_i}}]}$.
\newpage
\begin{center}
\begin{picture}(350,150)
\ArrowLine(70,140)(140,105)
\ArrowLine(140,105)(70,70)
\ArrowLine(210,105)(280,140)
\ArrowLine(280,70)(210,105)
\Gluon(140,105)(210,105){7}{7.5}
\end{picture}
\\{Fig.1: Tree-level Diagram}
\end{center}

\begin{center}
\begin{picture}(350,150)
\ArrowLine(70,140)(140,105)
\ArrowLine(140,105)(70,70)
\ArrowLine(210,105)(280,140)
\ArrowLine(280,70)(210,105)
\Gluon(140,105)(165,105){4}{4.5}
\GCirc(175,105){10.0}{0.5}
\Gluon(185,105)(210,105){4}{4.5}
\end{picture}
\\{Fig.2a: Gluon Self Energy due to  Particles}
\end{center}

\begin{center}
\begin{picture}(350,150)
\Line(70,140)(140,105)
\GCirc(105,122.5){5.0}{0.5}
\ArrowLine(140,105)(70,70)
\ArrowLine(210,105)(280,140)
\ArrowLine(280,70)(210,105)
\Gluon(140,105)(210,105){7}{7.5}
\end{picture}
\\{Fig.2b: Quark Self Energy due to SQCD Particles}
\end{center}

\begin{center}
\begin{picture}(350,150)
\ArrowLine(70,140)(140,105)
\Line(140,105)(70,70)
\GCirc(105,87.5){5.0}{0.5}
\ArrowLine(210,105)(280,140)
\ArrowLine(280,70)(210,105)
\Gluon(140,105)(210,105){7}{7.5}
\end{picture}
\\{Fig.2c Anti-Quark Self Energy due to SQCD Particles}
\end{center}

\begin{center}
\begin{picture}(350,150)
\ArrowLine(70,140)(140,105)
\ArrowLine(140,105)(70,70)
\Line(210,105)(280,140)
\GCirc(245,122.5){5.0}{0.5}
\ArrowLine(280,70)(210,105)
\Gluon(140,105)(210,105){7}{7.5}
\end{picture}
\\{Fig.2d Top Self Energy due to SQCD Particles}
\end{center}

\begin{center}
\begin{picture}(350,150)
\ArrowLine(70,140)(140,105)
\ArrowLine(140,105)(70,70)
\ArrowLine(210,105)(280,140)
\Line(280,70)(210,105)
\GCirc(245,87.5){5.0}{0.5}
\Gluon(140,105)(210,105){7}{7.5}
\end{picture}
\\{Fig.2e Anti-Top Self Energy due to SQCD Particles}
\end{center}

\begin{center}
\begin{picture}(350,150)
\Line(70,140)(140,105)\Line(140,105)(70,70)
\Gluon(140,105)(190,105){7}{7.5}
\DashLine(190,105)(225,140){1.5}
\DashLine(225,70)(190,105){1.5}
\Line(225,70)(225,140)
\Line(225,70)(280,70)\Line(225,140)(280,140)
\end{picture}
\\{Fig.3a: Triangle contribution from two stops and one gluino}
\end{center}

\begin{center}
\begin{picture}(350,150)
\Line(70,140)(140,140)\DashLine(140,140)(175,105){1.5}
\DashLine(175,105)(140,70){1.5}
\Line(140,70)(140,140)\Line(140,70)(70,70)
\Gluon(175,105)(225,105){7}{7.5}
\Line(225,105)(280,140)\Line(280,70)(225,105)
\end{picture}
\\{Fig.3b: Triangle contribution from two squarks and one gluino}
\end{center}

\begin{center}
\begin{picture}(350,150)
\Line(70,140)(140,105)\Line(140,105)(70,70)
\Gluon(140,105)(190,105){7}{7.5}
\Line(190,105)(225,140)
\Line(225,70)(190,105)
\DashLine(225,70)(225,140){1.5}
\Line(225,70)(280,70)\Line(225,140)(280,140)
\end{picture}
\\{Fig.3c: Triangle contribution from two gluinos and one stops}
\end{center}

\begin{center}
\begin{picture}(350,150)
\Line(70,140)(140,140)\Line(140,140)(175,105)\Line(175,105)(140,70)
\DashLine(140,70)(140,140){1.5}\Line(140,70)(70,70)
\Gluon(175,105)(225,105){7}{7.5}
\Line(225,105)(280,140)\Line(280,70)(225,105)
\end{picture}
\\{Fig.3d:Triangle contribution from two gluinos and one squark}
\end{center}

\begin{center}
\begin{picture}(350,150)
\ArrowLine(70,140)(280,140)
\ArrowLine(280,70)(70,70)
\DashLine(140,140)(140,70){1.5}
\DashLine(210,70)(210,140){1.5}
\end{picture}
\\{Fig.4a: Direct Box contribution from SQCD Particles}
\end{center}

\begin{center}
\begin{picture}(350,150)
\Line(70,140)(140,140)\Line(140,140)(210,70)\Line(210,140)(280,140)
\Line(280,70)(210,70)\Line(210,140)(140,70)\Line(140,70)(70,70)
\DashLine(140,140)(140,70){1.5}
\DashLine(210,70)(210,140){1.5}
\end{picture}
\\{Fig.4b:Crossed Box contribution from SQCD Particles}
\end{center}

\newpage
\begin{center}
\leavevmode\psfig{file=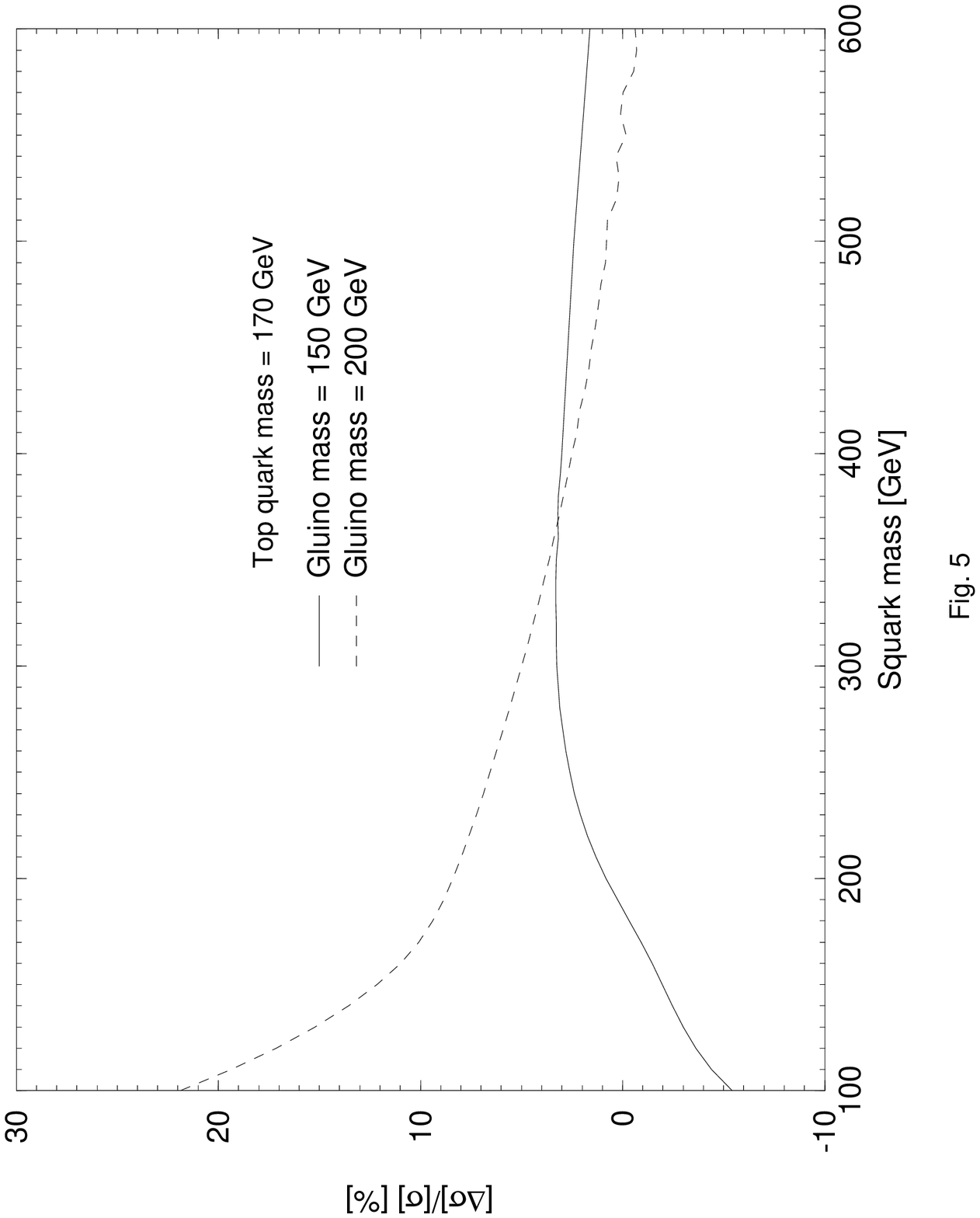,width=15cm,silent=0}
\end{center}

\newpage
\begin{center}
\leavevmode\psfig{file=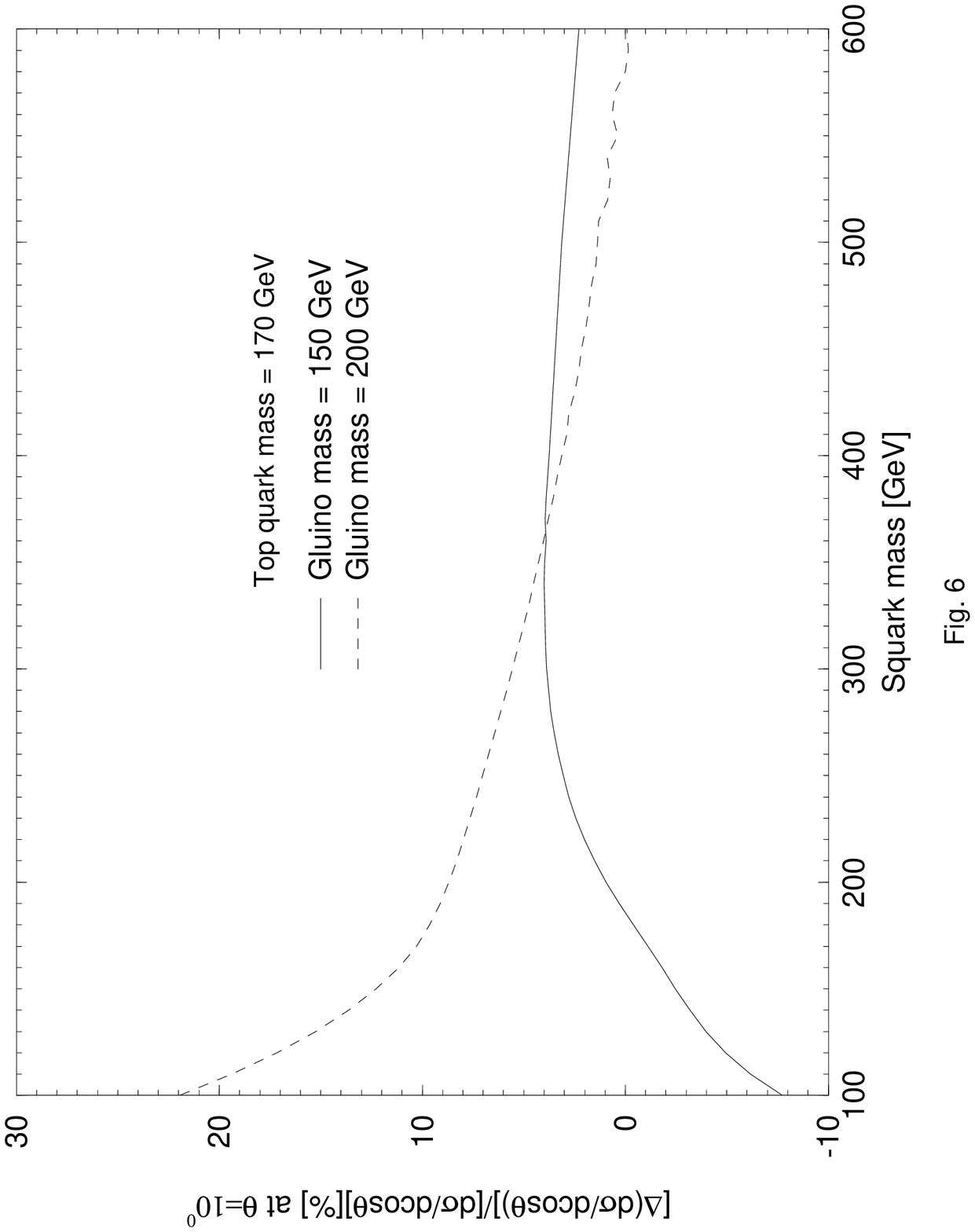,width=15cm,silent=0}
\end{center}

\newpage
\begin{center}
\leavevmode\psfig{file=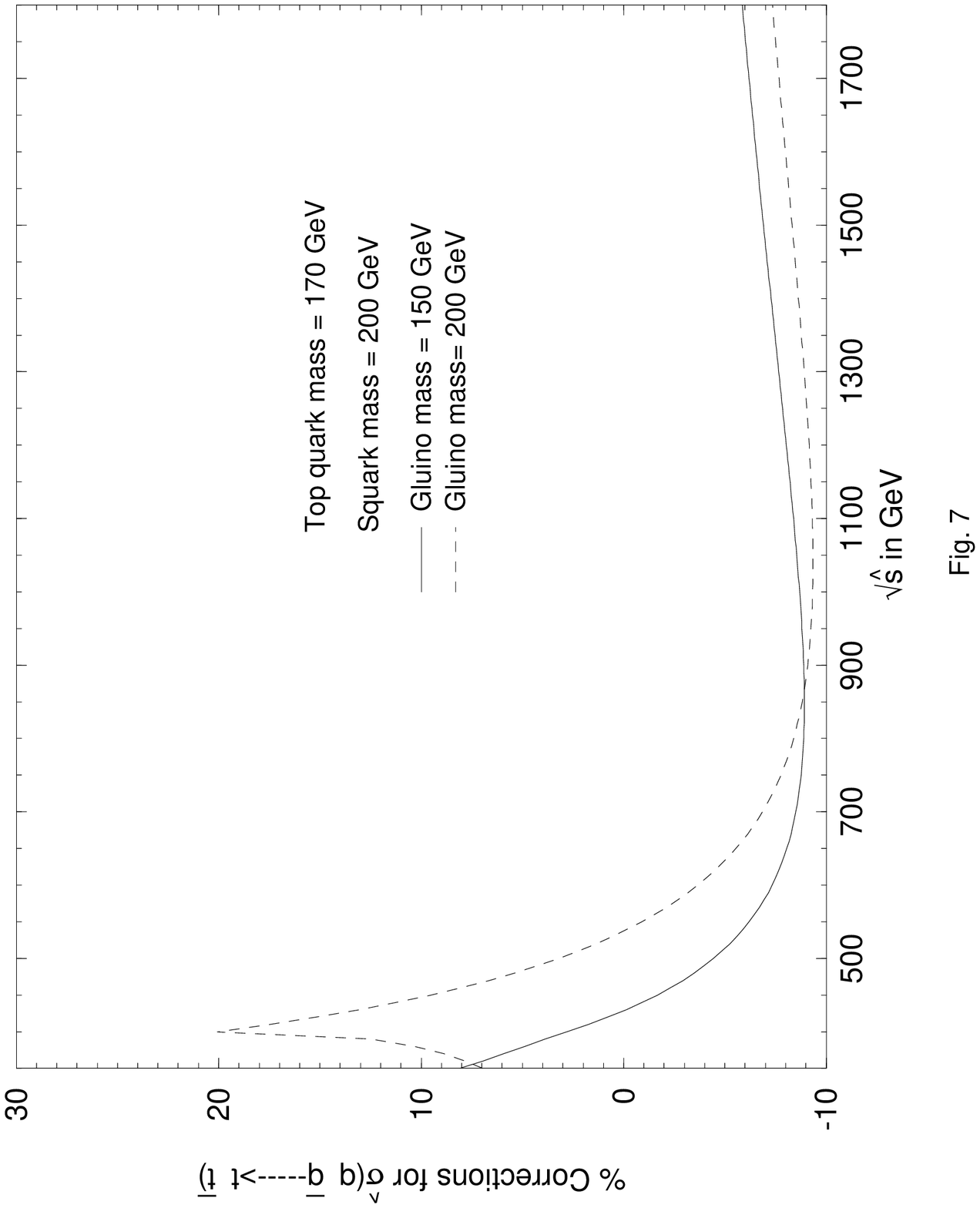,width=15cm,silent=0}
\end{center}

\newpage
\begin{center}
\leavevmode\psfig{file=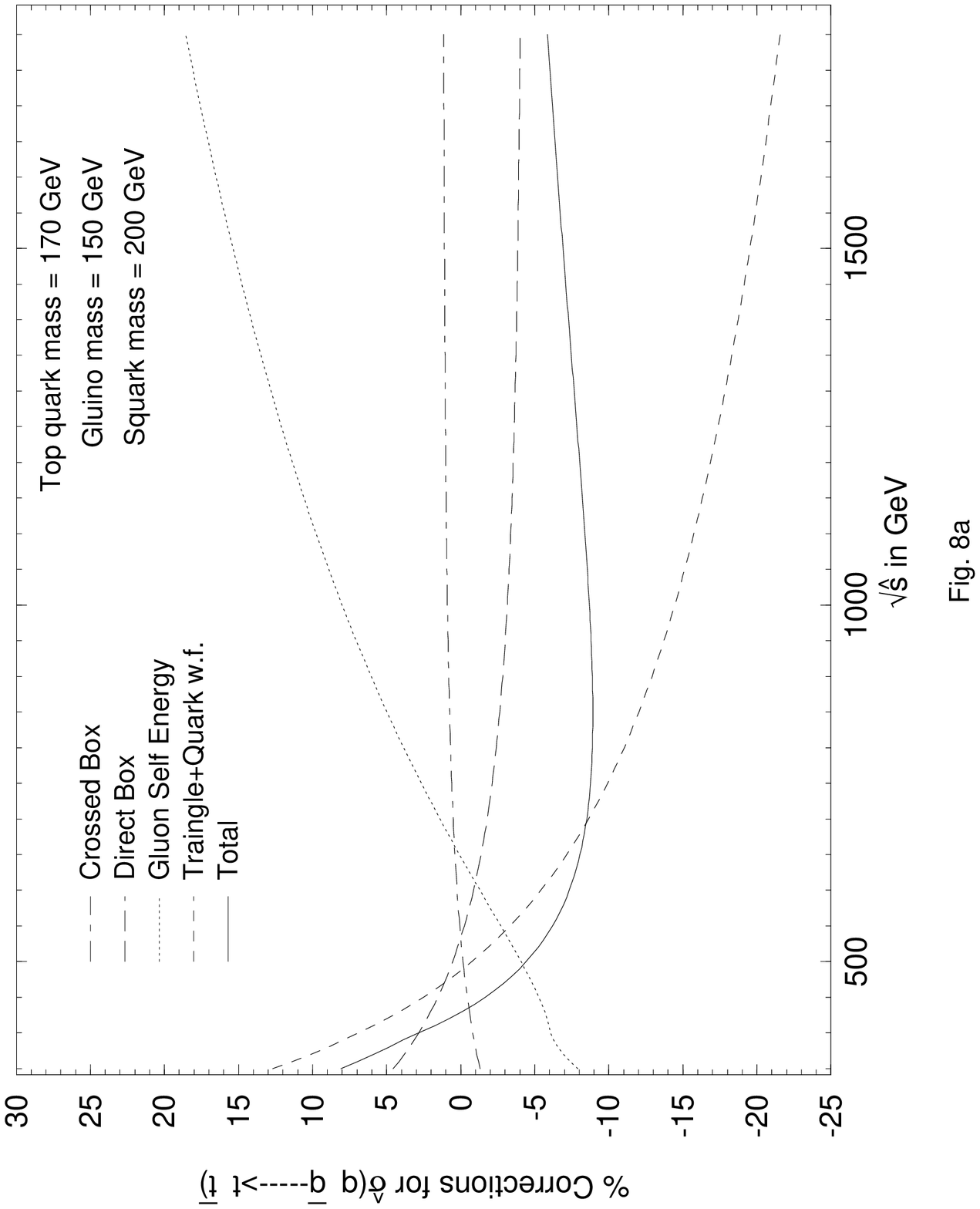,width=15cm,silent=0}
\end{center}

\newpage
\begin{center}
\leavevmode\psfig{file=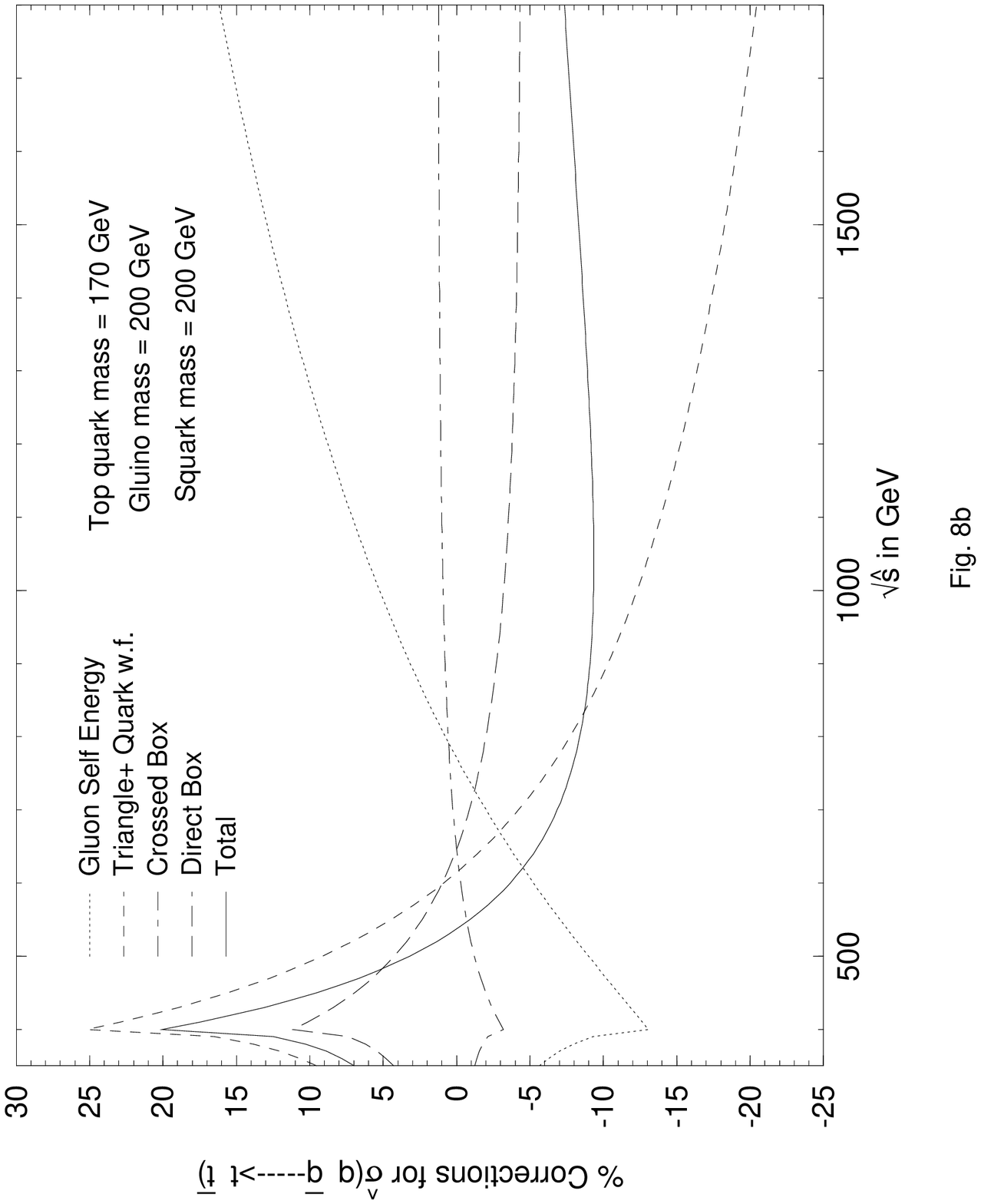,width=15cm,silent=0}
\end{center}

\newpage
\begin{center}
\leavevmode\psfig{file=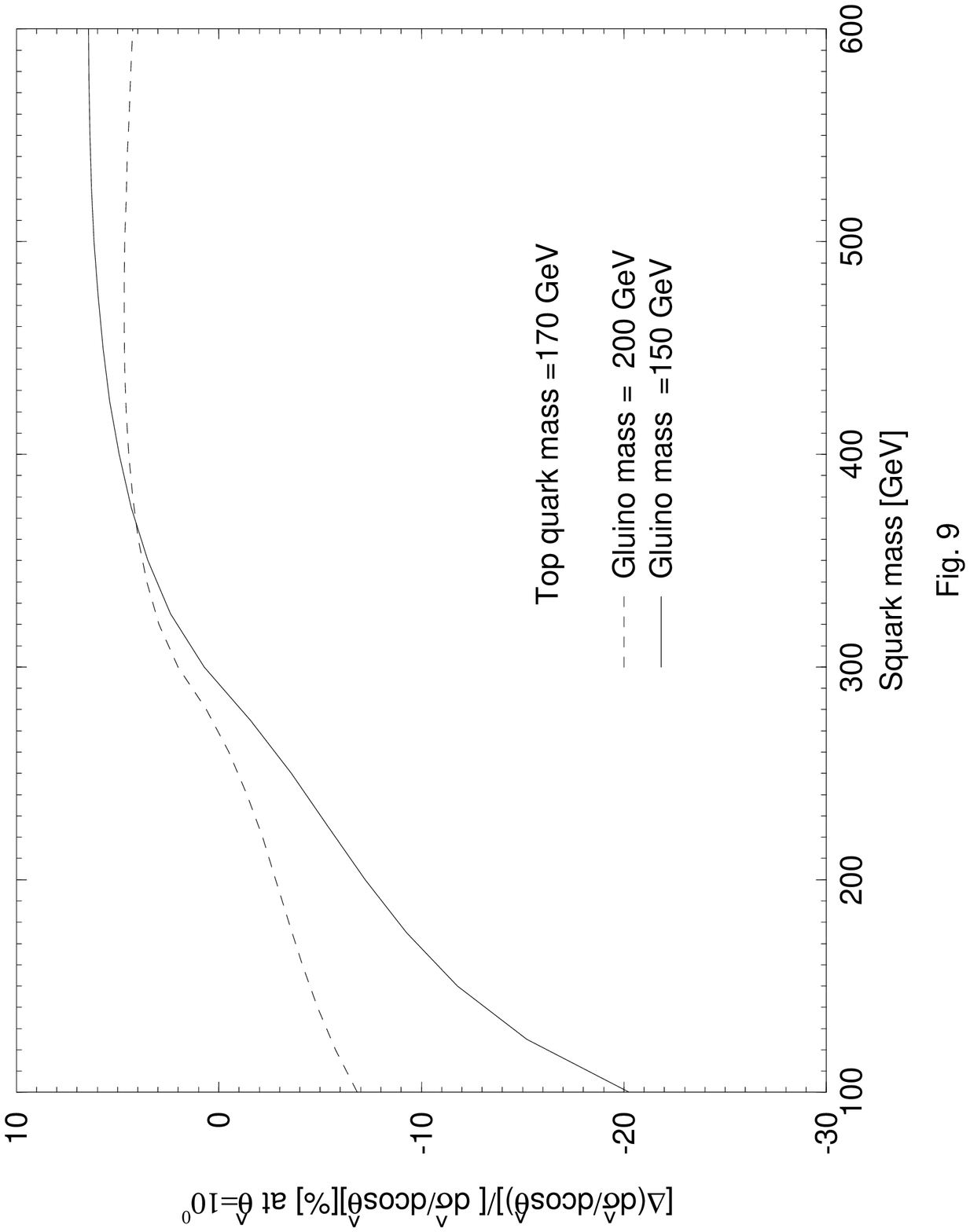,width=15cm,silent=0}
\end{center}

\end{document}